\documentclass{aa}
\usepackage[T1]{fontenc}
\usepackage{epsfig}
\usepackage{graphicx}
\usepackage{txfonts}
\usepackage{amssymb}
\usepackage{siunitx}
\usepackage{natbib} 
\usepackage{color}
\bibpunct{(}{)}{;}{a}{}{,}

\def\integ{{INTEGRAL\/}}
\def\isgri{{INTEGRAL/ISGRI}}

\def\xmm{{\em XMM-Newton}} 
\def\comptel{{COMPTEL}} 
\def\suzaku{{\em Suzaku}} 
\def\nustar{{\em NuSTAR}}
\def\rosat{{ROSAT}}
\def\hess{{\em H.E.S.S.}}
\def\ls{LS~5039 } 
\def\be{\begin{equation}} 
\def\ee{\end{equation}}

\def\f12{$\times$ $10^{-12}$ erg s$^{-1}$ cm$^{-2}$}
\def\lsim{\;\raise0.3ex\hbox{$<$\kern-0.75em\raise-1.1ex\hbox{$\sim$}}\;}
\def\gsim{\;\raise0.3ex\hbox{$>$\kern-0.75em\raise-1.1ex\hbox{$\sim$}}\;}

\begin{document} 

\title{Phase-resolved hard X-ray emission of the high-mass binary LS~5039:
a spectral hardening above 50~keV detected with \integ}
 
\author{M. Falanga\inst{1}
\and A. M. Bykov\inst{2}
\and Z. Li\inst{3,1}
\and{A. M. Krassilchtchikov}\inst{2}
\and{A. E. Petrov}\inst{2}
\and E. Bozzo\inst{4}
} 
 
\offprints{M. Falanga} 
 
\titlerunning{Phase-resolved studies of LS~5039}  
\authorrunning{M.~Falanga et al.}
  
\institute{International Space Science Institute (ISSI), Hallerstrasse 6, 3012 Bern,
Switzerland
\email{mfalanga@issibern.ch}   
\and Ioffe Institute, 26 Politechnicheskaya, 194021, St.~Petersburg, Russia
\and Key Laboratory of Stars and Interstellar Medium, Xiangtan University, Xiangtan 411105,
Hunan, China
\and ISDC Data Centre for Astrophysics, Chemin d'\'Ecogia 16, 1290 Versoix, Switzerland
} 
 
\date{} 
 
\abstract {} {\ls\ is an enigmatic high-mass gamma-ray binary which hosts 
a powerful O6.5V companion, but the nature of the compact object is still to be established using multi-wavelength observations.} 
{We analyzed phase-resolved multi-instrument spectra of nonthermal emission from \ls\  in order
to produce reliable spectral models, which can be further employed to select between various 
scenarios and theoretical models of the binary.} 
{The combined phase-resolved hard X-ray and MeV-range gamma-ray spectra obtained with \xmm, \suzaku,
\nustar, \integ, and \comptel\ indicate a meaningful spectral hardening above 50~keV. The spectral
break observed in both major phases of the binary may indicate the presence of an upturn in the spectrum of accelerated leptons which could originate from the interaction 
of wind from the O6.5V companion star  with the relativistic outflow from a yet unidentified compact object.} {}


\maketitle
 
\section{Introduction} 
\label{sec:intro}

\ls\ (also known as RX~J1826.2-1450) is a high-mass X-ray binary system in the constellation
of Scutum, whose high-energy emission was discovered within the \rosat\ Galactic Plane survey
\citep{Motch97}. The binary consists of an O6.5V companion star and a compact object
\citep{Casares05, Takata14}. 
The compact object, whose nature has not yet been established, is moving around the companion
star in an elliptic orbit with an eccentricity $e\approx$ 0.35 and an orbital period $P_{\rm
orb} = $3.9060 days \citep{Casares05, Aragona09, Sarty11}. The accuracy of the ephemeris first
established by \citet{Casares05}, was slightly improved by a combined multiband analysis by
\citet{chang16}.

Apart from the X-rays, extended (up to 1000 AU) radio emission \citep{Marti98}, MeV-range
\citep{Collmar14}, high-energy (HE) gamma-ray emission in the GeV range \citep{2000Sci...288.2340P,Abdo09}, and very
high-energy (VHE) gamma-ray emission \citep{Aharonian05} have also been 
detected from \ls. The extended radio emission is persistent and not orbitally modulated 
while the X-ray, MeV-range, and TeV-range emission from \ls\ show clear modulation with
orbital period. 
A modulation of GeV emission from \ls\ has also been found \citep{Abdo09,Hadasch12}, however this is in
anti-phase compared to the X-ray, MeV, and TeV light curves, which may indicate a different origin. 

Based on notable differences in the spectral energy distribution, \citet{Aharonian06} divided the orbit of \ls\ into two
phases: the superior conjunction phase (SUPC, $\phi \leq 0.45$ or $\phi \geq 0.9$, $\phi=0$
corresponding to the periastron) and the inferior conjunction phase (INFC, $0.45<\phi < 0.9$).
During the INFC, the VHE spectrum is well described by an exponential cutoff power law with
the index $\Gamma\sim1.85$ and the cutoff energy $E\sim8.7$ TeV, 
while during the SUPC, the spectrum is consistent with a steeper power law with
$\Gamma\sim2.53$. 
The nonthermal X-ray and MeV-range gamma-ray emission from \ls\ is likely of synchrotron
origin, while the TeV-range gamma-rays are likely produced via inverse Compton (IC) scattering
of the same distribution of accelerated leptons. However, the details and parameters of both
radiation mechanisms likely at work in \ls\ are still debatable 
\citep[see, e.g.,][and references therein]{Goldoni07, Takahashi09, Takata14}.   

\citet{2020arXiv200902075Y} studied the data on the 10--30 keV emission of \ls\, 
obtained with \nustar\ and \suzaku\ and found some signs of $\sim$ 9~s pulsations, perhaps indicating that the compact object in \ls\ is a neutron star. However, the pulsed fraction
inferred from these data vary from 0.68 down to less than 0.14 of the $\sim$ (8--11)\f12 flux
measured in the 10--30 keV band. In the 3--10 keV band, \nustar\ data have only allowed 
\citet{2020arXiv200902075Y} to set a $\sim$ 3\% upper limit on the pulsed fraction.
Another combined analysis of \nustar\ and \suzaku\ data by \citet{volkov21} also
revealed the 9~s pulsations, but the authors considered them not significant enough to
conclude on the presence of a neutron star in the binary. 

\citet{Hoffman09} analyzed about 3~Ms of \isgri\ exposures of \ls\ and
were able to show that its 25--200 keV light curve follows a similar orbital profile to the emission
observed in the 0.5--10 keV and TeV ranges. 
However, with such an exposure, the average 25--200 keV
INFC flux was estimated as (35.4$\pm$23.0)\f12 and for the SUPC flux only an upper limit of
14.5\f12 was set. Further analysis of \isgri\ data
was performed by \citet{chang16} with a multi-band comparison of different missions  light curves.
 
Below we analyze about 14 Ms of \isgri\ exposure of \ls\ 
(of which $\sim 9.5$ Ms of the effective exposure time are employed for the phase-resolved
spectral analysis) and model combined keV--MeV spectra
from data taken by  \xmm, \nustar, \suzaku,\  and \comptel. We also discuss a scenario whereby the hard emission of \ls\ originates from the zone where the powerful 
wind of the O6.5V companion collides with the outflow from the compact object.

It should be noted that despite the almost 14 Ms total exposure of \ls, the sensitivity of
\isgri\ is not enough to search for $\sim$ 10~s periods of 20--250 keV emission at the
$\sim10^{-11}$ erg s$^{-1}$ cm$^{-2}$ flux level. Hence, unlike the case with the Crab PWN, here it is not possible to distinguish the 20--250 keV emission coming from the \ls\ compact object (whose nature has not yet been established) from the emission generated in its surroundings where a pulsar wind or a fast jet from a black hole collides with the wind of its O6.5V companion.

\section{Observations and data reduction} 
\label{sec:obs}

The analyzed observations of \ls\ are briefly listed in Table~\ref{table:observations}, 
while a more detailed description is given in the following sections.

\begin{table}[h] 
{\small
\caption{Observations of \ls used in the present paper.}
\centering
\begin{tabular}{lcccc} 
\hline 
Mission & Obs. ID & Instrument & INFC  & SUPC  \\
 &  &  & ks & ks\\
\hline 
\noalign{\smallskip}  
\suzaku                   & 402015010   & XIS0\&XIS3    & 145 & 262  \\ 
                          &             & HXD    & 629 &  110\\

\xmm                 &  0151160201          & MOS2  &    10   &  --   \\
                     &  0742980101          & MOS2  &     --     &  72 \\

\integ               &  Revs. 50 -- 1939        & ISGRI  & 4252 & 5251 \\ 

\nustar             & 30201034002  & FPMA/B & 83 & 95 \\

{\it CGRO}$^a$             & 51 VPs 5--907  & \comptel & -- & -- \\
\hline  
\end{tabular}  
\tablefoot{ \tablefoottext{a}{For more details, we refer to \citet{Collmar14}, but 
the total effective exposure time was $\sim 7$~Ms.}}
\label{table:observations} 
}
\end{table} 

\subsection{INTEGRAL} 
\label{sec:integral}  

The International Gamma-Ray Astrophysics Laboratory  (\integ) observations are split into pointings with a typical duration of $\sim$ 2~ks \citep{w03}.
For the current analysis, we used all publicly available pointings performed in the direction of the source with position offset $\lesssim 12\fdg0$ from the center of the field of view, from the beginning of the \integ\ operations up to the time of the current analysis. These covered observations
from the satellite revolution 50 (March 12, 2003) to 1939 (April 8, 2018), giving a total exposure of 13.9~Ms (not corrected for instrumental effects at the source position). 
We only considered data collected with the IBIS/ISGRI coded mask instrument at
energies between 25 and 250 keV \citep{ubertini03,lebrun03}. All \integ\ data were analyzed
using version 10.2 of the OSA software distributed by the ISDC \citep{courvoisier03}. The algorithms used for the spatial and spectral analysis are described in \citet{gold03}. We first extracted a mosaic image by combining all available pointings in the 20--100~keV energy
band. In this mosaic, the source was detected at a significance of 26.5~$\sigma$, which is a vast improvement on the previous reported significance of 18.3~$\sigma$ and 7.7~$\sigma$ in the
(20--60 keV) as reported by \citet{chang16} and \citet{Hoffman09}, respectively.
We then extracted the IBIS/ISGRI light curve in the same energy band with the resolution of one
pointing and converted the arrival times to the \ls\ orbital phase using the ephemeris from
\citet{Casares05}. The light-curve data were subsequently separated into the two orbital phases
INFC and SUPC mentioned above (see Sect. \ref{sec:lc}). Concerning the spectral analysis, the
relatively low detection significance of the source in the long-term IBIS/ISGRI mosaic allowed
us to extract only one statistically meaningful spectrum for each of the INFC and SUPC orbital
phases (see Sec. \ref{sec:spec}). In order to maximize the signal-to-noise ratio of the data,
we adopted a response matrix created ad hoc  comprising 16 energy bins logarithmically spaced
between 20 and 250 keV. The total effective exposure times of the two spectra, corrected for all 
instrumental effects at the source position were of 5.251~Ms
for the SUPC phase and 4.252~Ms for the INFC phase.  

\subsection{XMM-Newton} 
\label{sec:xmm}  

Five observations of \ls\ have been recorded by \xmm, namely the Obs. IDs 0151160201 ($\phi\sim0.52-0.55$),
0151160301 ($\phi\sim0.53-0.56$), and 0202950201 ($\phi\sim0.46-0.51$) for the INFC phase, and  0202950301 ($\phi\sim0.99-0.02$) and 0742980101 ($\phi\sim0.93-0.14$) for the SUPC phase. The first four of them were analyzed by
\citet{Bosch05} and \citet{Kishishita09}. For the \xmm\ data, we analyzed all archived
observations in image mode from the European Photon Imaging Camera \citep[EPIC,][]{Jansen01,
Turner01,Struder01}. Following the standard data reduction threads introduced by the \xmm\
team,  the data reduction was carried out using the {\em Science Analysis System} 18.0.0. The
source spectra were extracted from a circular region with a radius of $40\arcsec$ for both the
pn and MOS, while the background spectra were extracted from a source-free circular region with
a radius of $100\arcsec$ (MOS) and  $50\arcsec$ (pn). For each observation, the response (RMFs) and
ancillary files (ARFs) were produced by {\tt rmfgen} and {\tt arfgen}, respectively. All spectra can
be fitted by an absorbed power-law model, with the power-law index and the unabsorbed flux
(1-10 keV) in the range 1.3--1.6 and $6\times10^{-12}-1\times10^{-11}$ erg s$^{-1}$ cm$^{-2}$,
respectively. In the energy range 1--10 keV, the flux in Obs. ID 0202950301 is $\sim 6\times10^{-12}~{\rm erg~s^{-1}~cm^{-2}}$, which is 12\% lower than the averaged flux in the SUPC phase observed by \suzaku.  Moreover, we find that  the Obs. ID 0151160201 (0742980101) shows
a similar power-law index and flux to the averaged {\em{Suzaku}} spectrum in the INFC (SUPC) phase. Therefore, we only use the MOS2 spectrum from Obs. ID 0151160201 for the INFC phase. For the SUPC phase, we extracted the MOS2 spectrum from Obs. ID 0742980101, which has never been
published before.

\subsection{SUZAKU} 
\label{sec:suzaku}  

\ls\ was observed with {\em{Suzaku}} \citep{Serlemitsos07}  from September 9 to 15, 2007, for
$\sim$ 200 ks covering one and half orbital periods.  We followed the data-reduction processes
from \citet{Takahashi09}, who reported the spectral results. 
For the XIS data analysis, the cleaned events were obtained using the ftools
command {\tt aepipeline}. We separately extracted the source spectra in INFC and SUPC phases from a circular region with a radius of 3\arcmin. The corresponding background spectra were extracted in a source-free region with the same radius. We combined the spectra from the XIS0 and XIS3 to increase the source photons. The RMF and ARF files were generated using {\tt xisrmfge} and {\tt xissimarfgen}, respectively.
The standard criteria are applied to screen the Hard X-ray Detector  \citep[HXD,][]{Takahashi07} data using the ftools
command {\tt aepipeline}. The time-averaged spectra in INFC and SUPC phases of \ls\ were extracted. The HXD-PIN background
is dominated by three components, namely  the time-variable instrumental background (nonX-ray
background (NXB)), the cosmic X-ray background (CXB), and the Galactic ridge X-ray emission
(GRXE). We used similar methods to those introduced by \citet{Takahashi09} to obtain all the
background spectra. However, we chose an updated RMF file ae\_hxd\_pinhxnome4\_20080129.

We used an absorbed power-law model to fit the joint \suzaku/XIS-HXD spectra, and found that $N_{\rm H}$ are $0.74\pm0.03$ and $0.76\pm0.03$ in units of $10^{22} \,{\rm cm}^{-2}$, and the power-law index is $1.51\pm0.02$ and $1.60\pm0.02$ for the INFC and SUPC phases, respectively. The results are consistent with the values reported in \citet{Takahashi09} at 1$\sigma$ confidence level.

\subsection{NuSTAR}
\label{sec:nustar}

\nustar\ observed \ls\ on September 1, 2016 (Obs ID. 30201034002). We analyzed the data for
SUPC and INFC separately.  We cleaned the event file using the \nustar\ pipeline tool {\tt
nupipeline} for both FPMA and FPMB by activating the {\tt usrgtifile} option. The spectra of
SUPC and INFC were extracted from a circular region with a radius of  $60\arcsec$ centered on
the source location of \ls using {\tt nuproducts}, and the response files were produced
simultaneously. To extract the background spectra, we chose a source-free circular background
region located on the same chip with a radial aperture of $60\arcsec$. Finally, we obtained
the spectra during SUPC and INFC phases with the exposure time of 95.23 ks and 82.86 ks,
respectively. We used the spectra from FPMA and FPMB to perform spectral analysis.

\subsection{COMPTEL}
\label{sec:comptel}

\ls\ was observed with the imaging Compton telescope \citep[\comptel,][]{Schoenfelder93} for a total effective exposure time of around 7 Ms with the first results reported in \citet[][]{Collmar14}. They extracted the flux in the standard \comptel\ energy bands 1--3, 3--10, and 10--30 MeV and subdivided into two orbital phases -- INFC and SUPC. We refer to \citet[][]{Collmar14} for more details. Therefore, the data used for our spectral analysis (see Sect. \ref{sec:spec}) were taken from \citet[][see Table 5]{Collmar14}. In order to perform the XSPEC spectral fitting with the \comptel\ data, we used the ftools command {\tt flx2xsp} to convert the flux to spectra and response files for the INFC and SUPC orbital phases.

\section{Orbital period modulation} 
\label{sec:lc} 

\begin{figure}[h] 
\centering 
\centerline{\epsfig{file=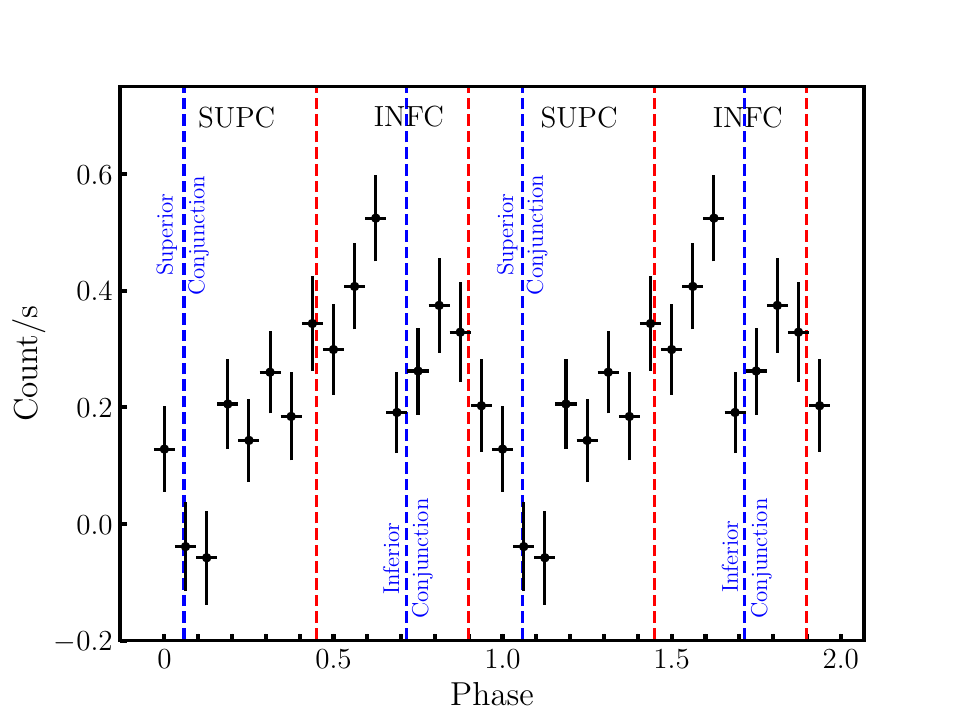,scale=0.6}} 
\caption{\isgri\ (20--100 keV) orbital profile of \ls. The light curve has been folded into 16
phase bins at the orbital period of 3.90603 days and epoch time T$_0$ = 51942.59 MJD (at
orbital phase $\phi = 0.0$). The red dashed lines indicate the interval of the INFC and SUPC
orbital phases. The superior conjunction and inferior conjunction are shown as blue dashed
lines: $\phi = 0.058$ and $\phi = 0.716$, respectively. The  orbital periastron and apastron
are at the phases $\phi = 0.0$ and $\phi = 0.5$, respectively. The period is repeated once for
clarity.}
\label{fig:fig1} 
\end{figure}

The \isgri\ lightcurve, 20--100 keV band, has been extracted over all available pointing
(6909) of roughly $\sim$ 2 ks each, and folded into 16 orbital phase bins using the ephemeris 
derived by \citet{Casares05}. The orbital light curve is shown in Fig.~\ref{fig:fig1}.
Please note that, as our observation starts 767 days (March 12, 2003) and ends 6273 days (April
8, 2018) after the considered epoch times, $T_0 = 51942.59 \pm 0.1$ MJD\footnote{The epoch is
given at orbital phase $\phi=0$} (February 2, 2001), with the given orbital period $P_{\rm
orb} = 3.9063 \pm 0.00017$ days, possible phase uncertainties of $\sim 0.03$
and $\sim 0.27,$ caused by the orbital error will be introduced. However, scanning the epoch time by some mean phase shift values between 0.03 and 0.27 did not improve the orbital period modulation amplitude, and as our orbital time bin is one-sixteenth of the orbital period,
consideration of the phase correction proposed by \citet{chang16} did not significantly improve
the shape (the maximal amplitude) of the folded light curves. 

If we compare  Fig.~\ref{fig:fig1} with previously reported \isgri\ folded light curves \citep{Hoffman09,chang16}, the orbital profile is modulated at the same amplitude and shows
a single peaked profile with its maximum and minimum at phase $ \phi \sim 0.65$ and $\phi \sim
1.2$, respectively. The  profile is also comparable to the \suzaku\ 1--10 keV, 15--40 keV,
\nustar\ 3--60 keV, \comptel\ 10--30 MeV, {\sl Fermi} 30-70 MeV and 70--100 MeV, and \hess\ $<$
1 TeV reported orbital light curves
\citep{Takahashi09,Kishishita09,volkov21,Collmar14,Aharonian06}. Please note that the Fermi 0.1--3 GeV light curves show anti-correlation compared with the above-mentioned orbital profiles \citep[see e.g.,][]{chang16}. Starting from the
superior conjunction phase, the orbital profile increases to its maximum, $\phi \sim 0.6$, then
suddenly decreases slightly near the inferior conjunction phase, before increasing again to
a second smaller maximum and then decreasing to its minimum at the superior conjunction orbital
phase (see Fig.~\ref{fig:fig1}). It is clear that the orbital profile is affected at the
inferior or superior conjunction phases rather than at the orbital periastron and apastron
passage. 

\section{Spectral analysis} 
\label{sec:spec}

The spectral analysis was carried out using XSPEC version 12.6 \citep{arnaud96}. All
uncertainties in the spectral parameters are given at 1$\sigma$ confidence level for a
single parameter. We divided the data into two orbital phases, namely SUPC ($\phi \leq 0.45$ or
$\phi \geq 0.9$, periastron at $\phi=0$) and INFC ($0.45<\phi < 0.9$), respectively. The
spectra from \xmm/\suzaku/\nustar/\integ\ were grouped to ensure each channel has 25 photons
at least. The \comptel\ spectra were converted from the measured flux as introduced in
Sect.~\ref{sec:comptel}. 

We select the energy ranges 0.5--10 keV, 1--10 keV, 3--70 keV, 15--60 keV, 25--250 keV, 1--30 MeV for \xmm, \suzaku/XIS, \nustar/FPMA/FPMB, \suzaku/HXD, \integ/ISGRI and \comptel, respectively. We first fit the joint broad-band spectra with a simple model consisting of a photoelectrically absorbed power-law model, {\tt phabs*powerlaw}. 
A multiplication factor is included in the fit to take into account the uncertainties in the cross-calibration of the instruments, and it is fixed at unity for the \nustar/FPMA data because it is well calibrated and its energy range overlaps with all other spectra, except for the \comptel\ data. The multiplication factor is a free parameter for all other spectra. If the \comptel\ data are not taken into account, the power-law model fits the spectra well, and the indices are $1.59\pm0.01$ and $1.59\pm0.01$, with a red-$\chi^2$/d.o.f. of 0.94/2503 and 1.01/2941 for INFC and SUPC phases, respectively. For the INFC phase, the multiplication factor of \xmm\ and \suzaku/XIS are 15\% lower than \nustar. For the SUPC phase, all multiplication factors are close to one, except the multiplication factor of \suzaku/HXD, which is $1.84\pm0.13$. The results are comparable with the other works \citep[see, e.g.,][]{2020arXiv200902075Y,volkov21,Yoneda2021}, while the residuals of the \integ/ISGRI data show an excess around 50 keV. When the \comptel\ data are considered, the power-law indices are not changed, but the best-fit multiplication factor of the \comptel\ spectra is $3.14\pm0.42$ and $3.34\pm0.66$ for INFC and SUPC phases, respectively. However, if we assume that  the available \comptel\ data are well calibrated and fix its multiplication factor at unity, then the single power-law model cannot describe the data despite the red-$\chi^2$ being close to one. This is because the whole keV-MeV data set is dominated by the data at low energies. Therefore, the high-energy data of \integ/ISGRI and \comptel\  indicate departure from a single power-law model at sufficiently high photon energies. This finding was also noticed by \citet[][see Fig. 11]{Yoneda2021} using the \nustar/\comptel\ data, suggesting an additional harder spectral component.

\begin{figure*}[h]
\centering
\hbox{
\psfig{figure=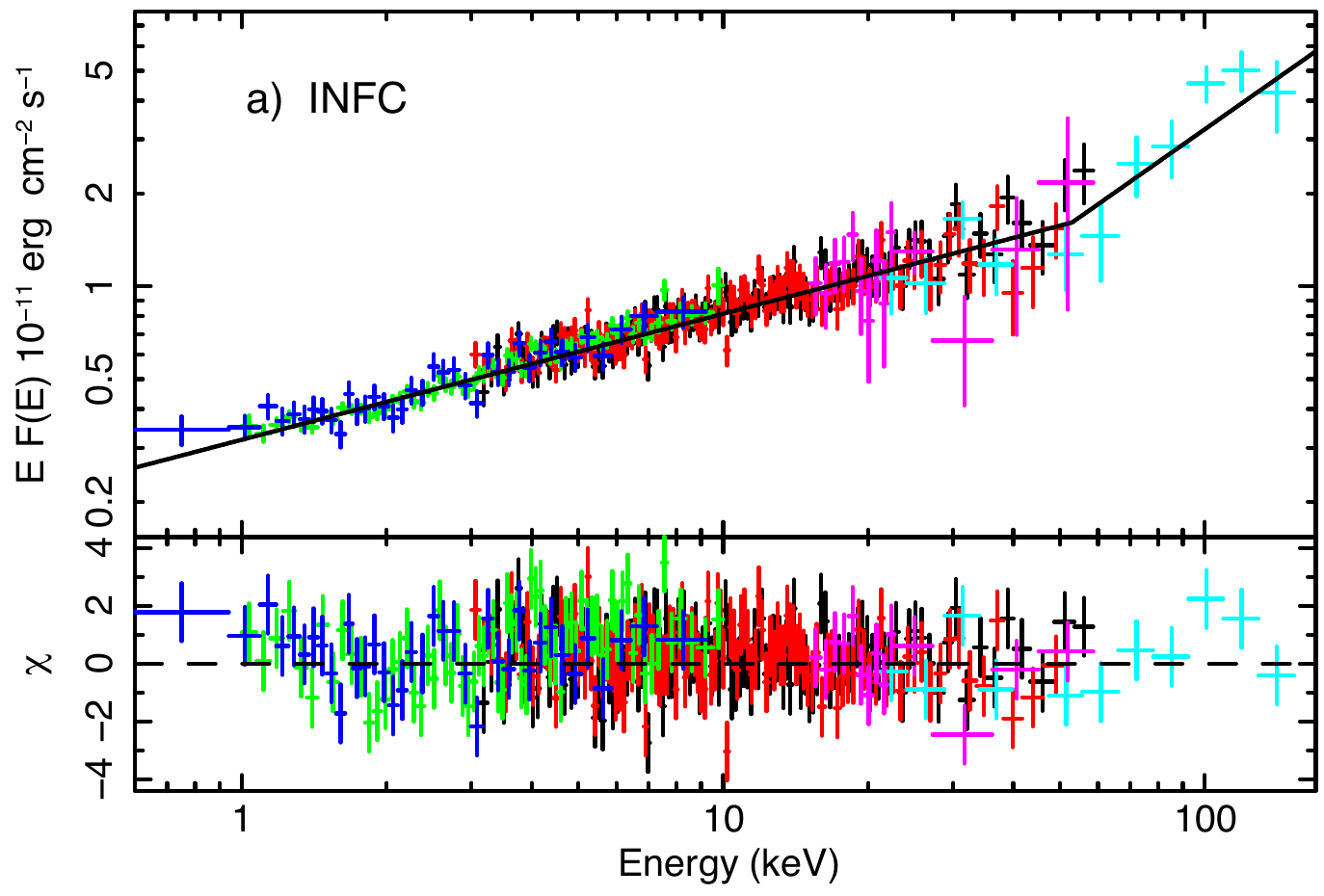, angle=-90, scale=0.36}
\hspace{0.35cm}
\psfig{figure=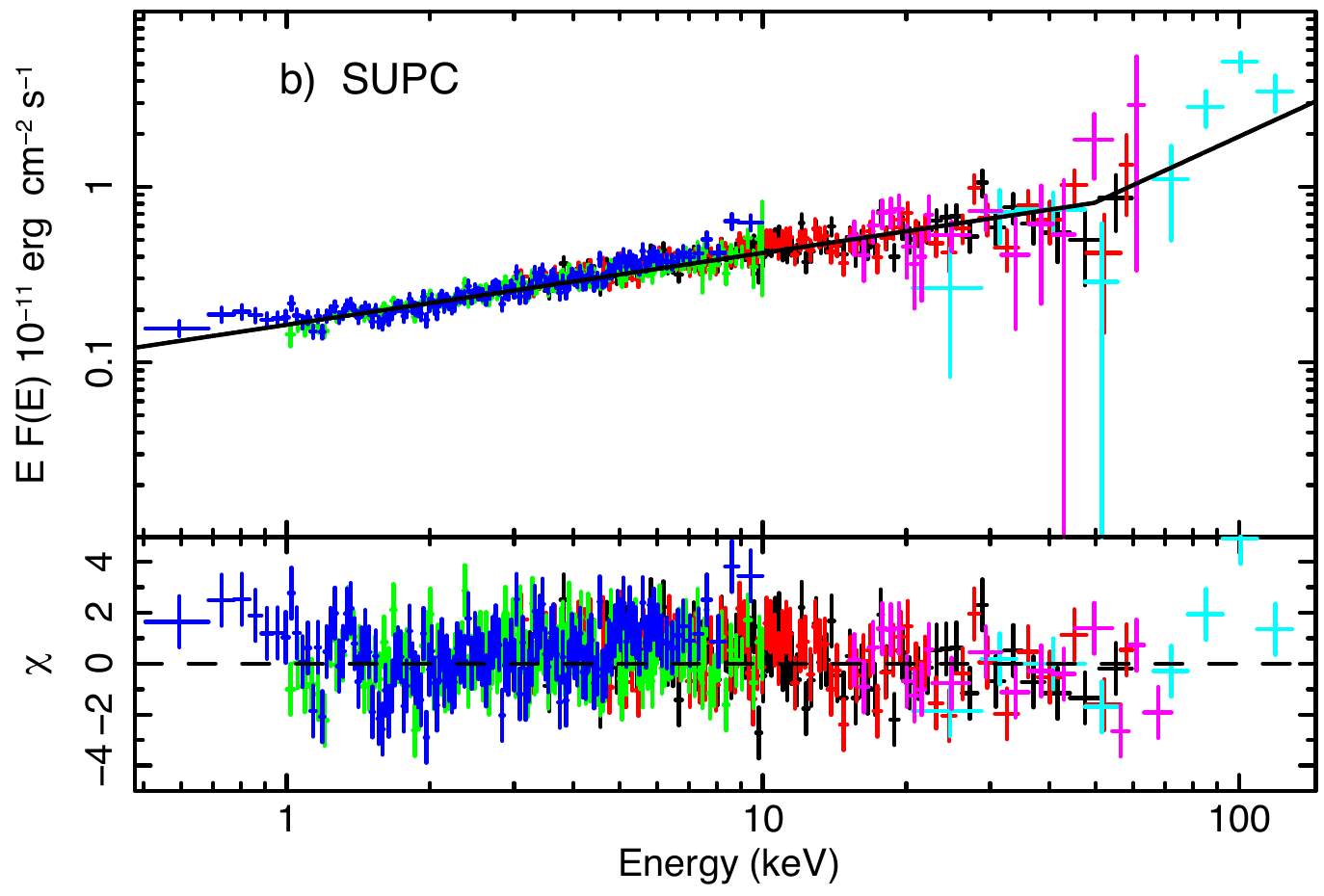, angle=-90, scale=0.36}
}
\vspace{0.15cm}
\hbox{
\psfig{figure=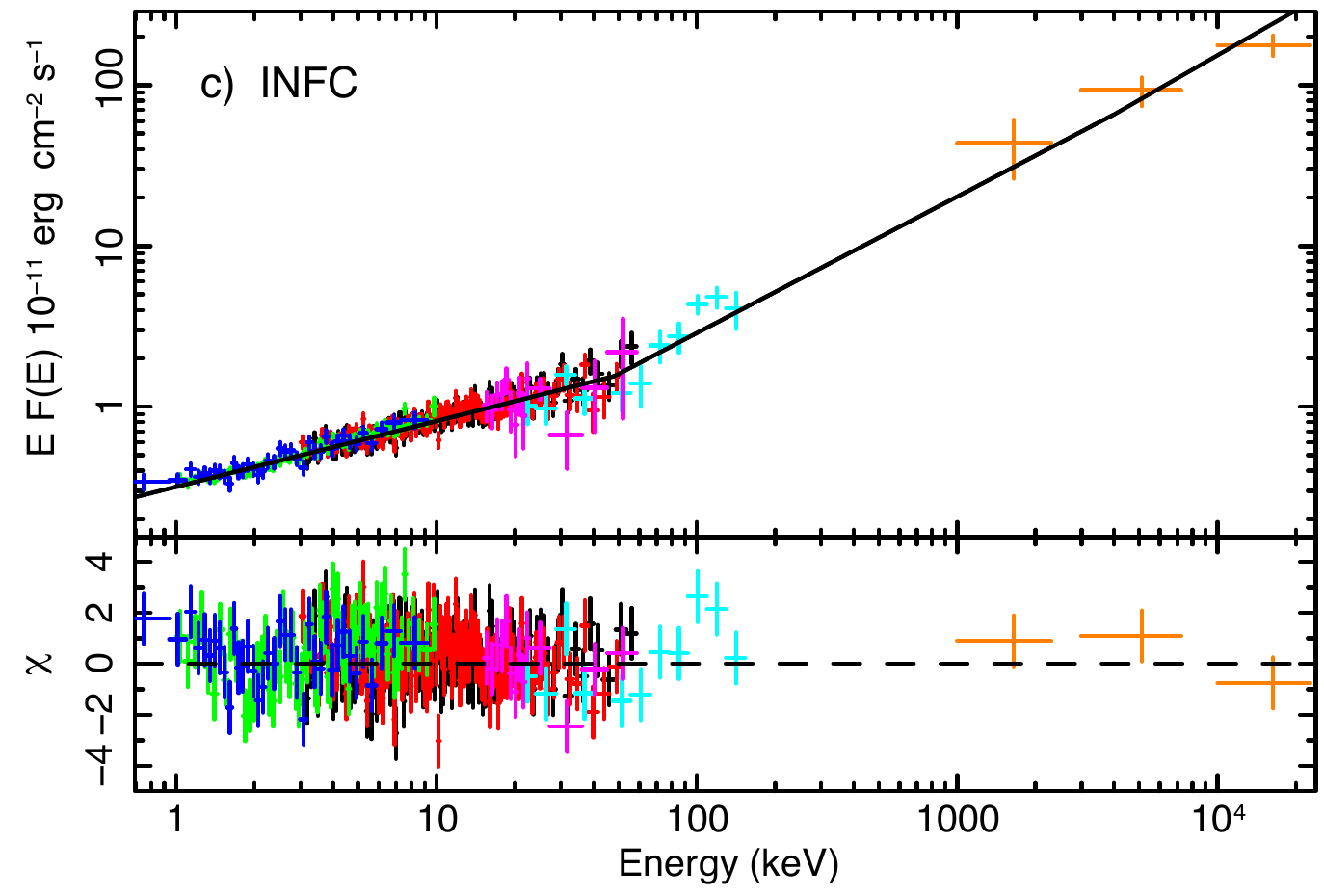, angle=-90, scale=0.36}
\hspace{0.35cm}
\psfig{figure=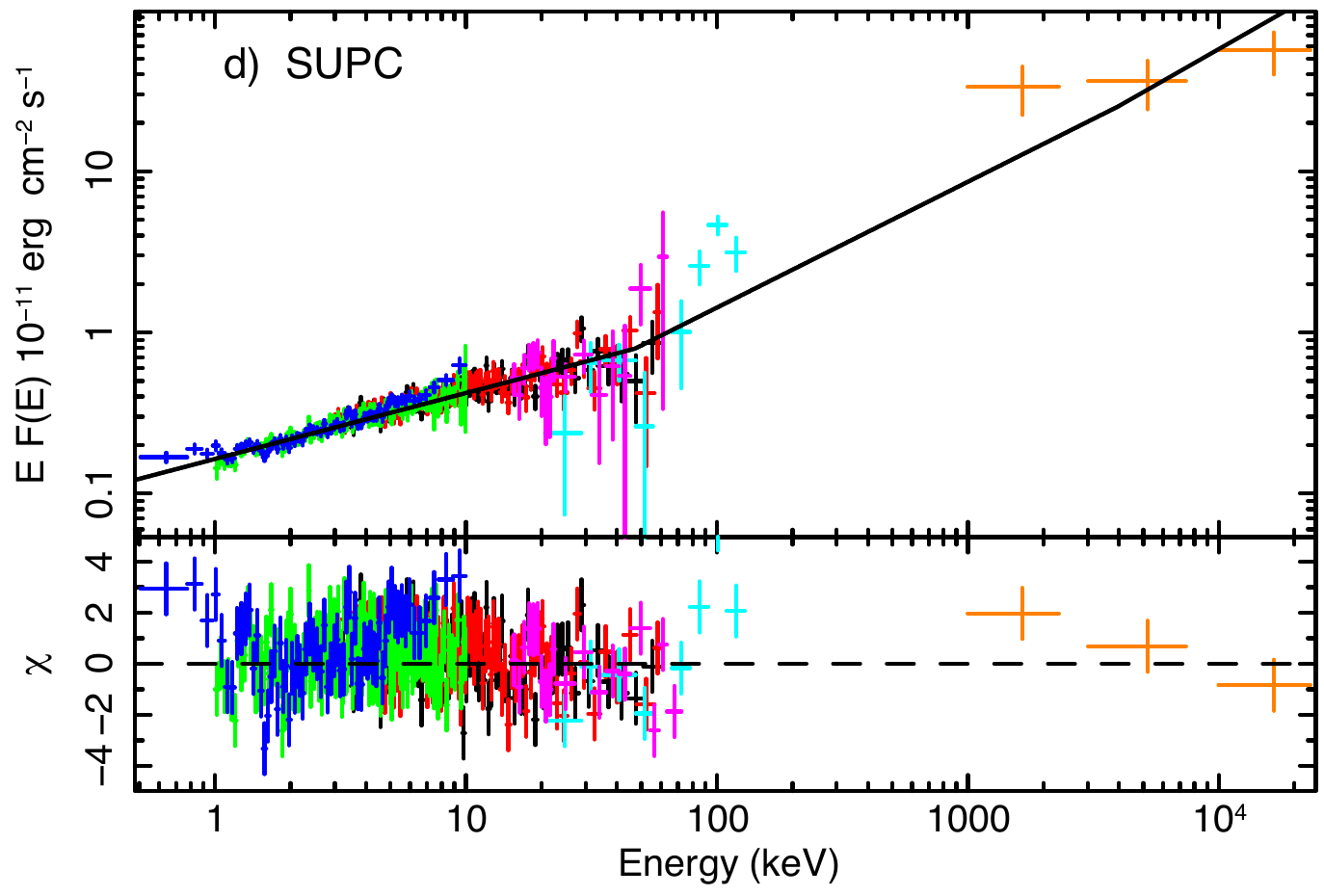, angle=-90, scale=0.36}
}
\vspace{0.15cm}
\hbox{
\psfig{figure=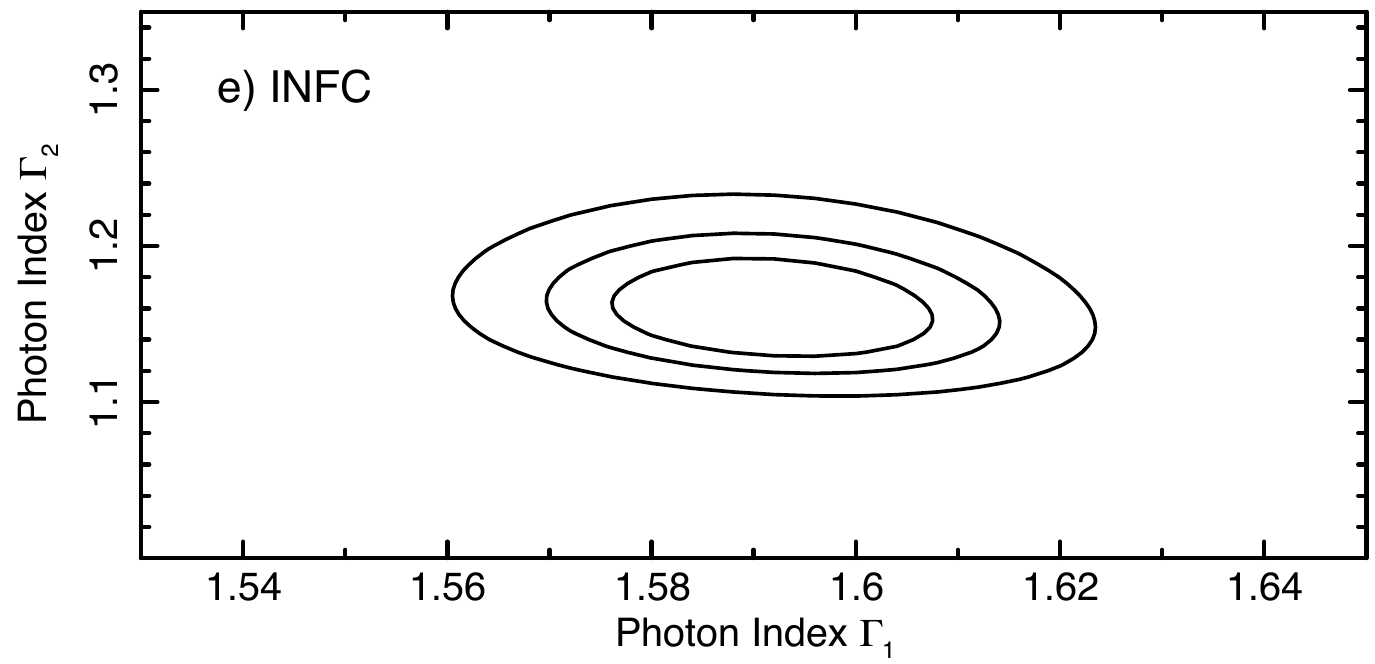, angle=-90, scale=0.36}
\hspace{0.35cm}
\psfig{figure=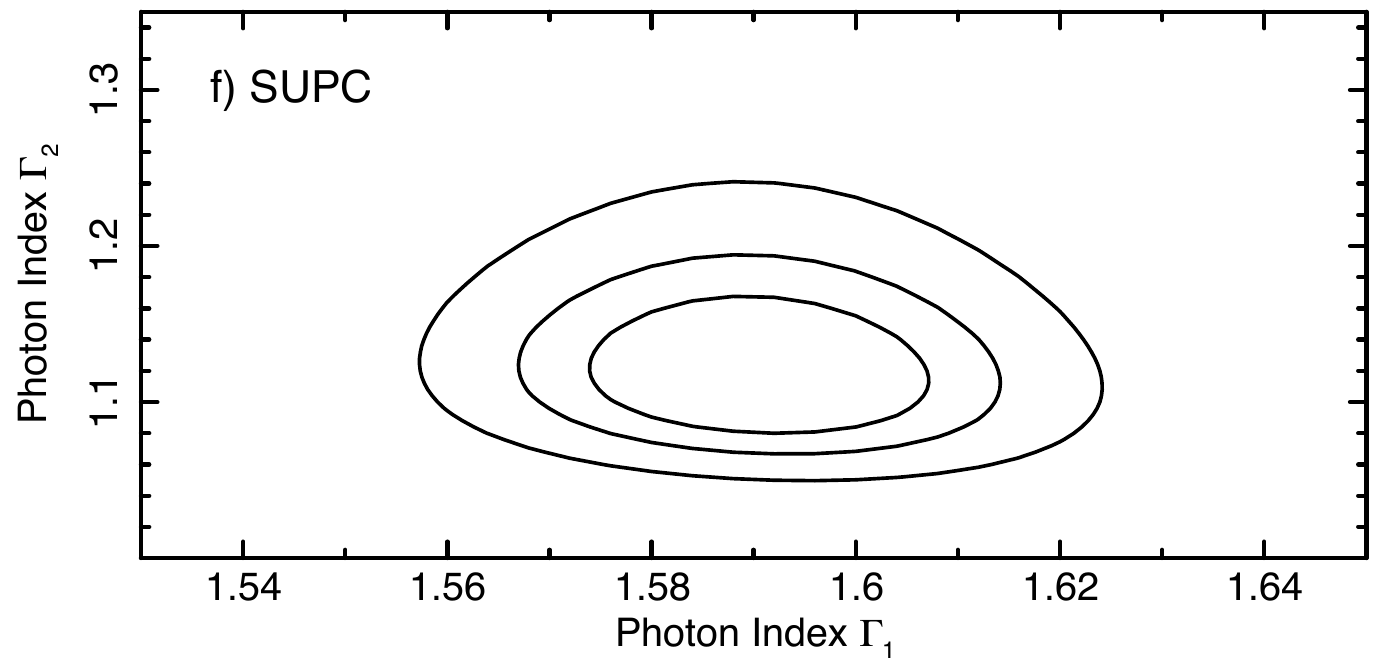, angle=-90, scale=0.36}
}
\caption{(a) and (b) are the unfolded absorbed broad-band spectra of \ls\ in the 0.5
-- 250 keV energy range and separated in the INFC and SUPC phases. (c) and (d) include the \comptel\ data and thus cover the 0.5 -- 2$\times10^4$ keV energy
range. The data points are obtained from the \xmm/MOS (blue points, 0.5--9 keV), the
\nustar/FPMA/FPMB (black/red points, 3--79 keV) \suzaku/XIS/HXD (green and magenta points, 1--10 keV and
10--70 keV, respectively), \integ/ISGRI (cyan points, 24--250 keV), and \comptel\ (orange
points, 1--30 MeV). The fit  represented in the top
panels with a solid line was obtained with the {\tt bknpow} model. The residuals from the best fits are shown in the lower panels. (e) and (f) are the contours corresponding to the spectral fit (c)
and (d) of the broken power-law indices separated in the INFC and SUPC phases,
respectively.}
\label{fig:spec_eeuf}
\end{figure*}

We then replaced the simple power law with a broken power-law model, {\tt phabs*bknpower}, to make a best fit of the broad-band spectra. The main parameters of the model are the hydrogen column density, $N_{\rm H}$, the break energy, $E_{\rm cut}$, and the photon indices, $\Gamma_1$ and $\Gamma_2$, below and above $E_{\rm cut}$. In this case, a multiplication factor was also included to account for the uncertainties in the cross-calibration of the instruments, where it was fixed at unity for \nustar/FPMA data. First, we only
consider the spectra in the energy range 0.5--250 keV. The best-fit parameters are reported in
Table~\ref{tab:table2}. We then extend the energy range up to 30 MeV by including the
\comptel\ spectra, and show the best-fit parameters in Table~\ref{tab:table3}. We note that the  multiplication factors of the \comptel\ are now around one-third lower for both INFC and SUPC phases compared with \nustar/FPMA spectra. If we fix the multiplication factor of \comptel\ data at one, the broken power-law model can also describe the broad-band spectra up to MeV with a slightly larger and acceptable red-$\chi^2$.  The unabsorbed bolometric fluxes are determined by the convolution model {\tt cflux} in the 0.5--250 keV and
0.5--2$\times10^4$ keV energy ranges.  The interstellar absorption column density is
consistent with previous results. Figure~\ref{fig:spec_eeuf} (a-d) shows the unfolded
spectra and the best-fit models. 

For the broad-band 0.5--250 keV spectra, we find the break energy, $E_{\rm cut}\sim50~{\rm
keV}$,  and the power-law indices, $\Gamma_1\sim1.6$ and $\Gamma_2\sim0.9$, consistent within the errors
between the SUPC and INFC phases. For the broad-band 0.5--2$\times10^4$ keV spectra, we
obtain a larger index, $\Gamma_2\sim1.15$, while other parameters, $N_{\rm H}$, $E_{\rm cut}$,
and $\Gamma_1$ are highly consistent at 1$\sigma$ confidence level compared with the results
from the broad-band 0.5--250 keV spectra. The contour plots between $\Gamma_1$ and $\Gamma_2$
are shown in Fig.~\ref{fig:spec_eeuf} (e) and (f).

\begin{table}[h] 
\caption{\label{table:spec} Spectral fitting to the broadband spectra in INFC and SUPC phases of \ls\  with the \xmm, \suzaku, \nustar, and \integ\, observations.}
\centering
\begin{tabular}{lll} 
\hline 
{\sc phabs$\times$bknpowe} &  INFC & SUPC\\
\hline 
$N_{\rm H} (10^{22} {\rm cm}^{-2})$ & $0.81\pm 0.02$ & $ 0.72\pm 0.01$\\ 
$E_{\rm cut}$ (keV) & $54.9\pm 7.4$ & $49.6\pm 5$\\ 
$\Gamma_{\rm1}$ & $1.59\pm0.01$ & $1.59\pm 0.01$\\ 
$\Gamma_{\rm2}$ & $0.88\pm0.15$ & $0.76\pm 0.21$\\ 
$C_{NuSTAR {\rm /FPMA}}$ & 1 (fixed) & 1 (fixed)\\
$C_{NuSTAR /FPMB}$ & $0.99\pm 0.01$ & $0.99\pm 0.02$\\
$C_{Suzaku /XIS}$ & $0.85\pm 0.01$ & $1.01\pm 0.02$\\
$C_{Xmm-Newton}$ & $0.83\pm 0.02$ & $0.99\pm 0.02$\\
$C_{\rm \integ /ISGRI}$ & $1.05\pm 0.08$ & $0.83\pm 0.12$\\
$C_{Suzaku /HXD}$ & $1.33\pm 0.09$ & $1.83\pm 0.13$\\
$\chi^{2}/{\rm d.o.f.}$ & 2322/2499 & 2945/2939\\
$F_{\rm bol}$\tablefootmark{a} ($10^{-11}$ erg s$^{-1}$ cm$^{-2}$) & $6.64\pm0.33$ & $2.68\pm0.16$\\
\hline  
\end{tabular}  
\tablefoot{Best parameters determined from the fits to the average broad-band spectrum of \ls\  performed with the \xmm, \suzaku, \nustar, and \integ\, data, using the
absorbed two-segment broken power-law (i.e., with one break energy) model. The  multiplication factor  is provided for all instruments. \\
\tablefoottext{a}{Unabsorbed flux in the 0.5--250 keV energy range.}}
\label{tab:table2} 
\end{table}

\begin{table}[h] 
\caption{\label{table:spec} Spectral fitting to the broadband spectra in INFC and SUPC phases of \ls\  with the \xmm, \suzaku, \nustar, \integ,  and \comptel\, observations.}
\centering
\begin{tabular}{lll} 
\hline 
{\sc phabs$\times$bknpowe} &  INFC & SUPC\\
\hline 
$N_{\rm H} (10^{22} {\rm cm}^{-2})$ & $0.81\pm0.02$ & $0.72\pm0.01$\\ 
$E_{\rm cut}$ (keV) & $47.4^{+8.7}_{-4.5}$ & $47.1\pm8.1$\\ 
$\Gamma_{\rm 1}$ & $1.59\pm 0.01$ & $1.59\pm 0.01$\\ 
$\Gamma_{\rm 2}$ & $1.16\pm 0.15$ & $1.14\pm0.25$\\ 
$C_{NuSTAR/FPMA}$ & 1 (fixed) & 1 (fixed)\\
$C_{NuSTAR/FPMB}$ & $0.99\pm0.02$ & $0.99\pm0.02$\\
$C_{Suzaku/XIS}$ & $0.85\pm0.01$ & $1.01\pm0.02$\\
$C_{Xmm-Newton}$ & $0.83\pm0.02$ & $0.99\pm 0.02$\\
$C_{\rm \integ/ISGRI}$ & $1.09\pm 0.08$ & $0.92\pm 0.10$\\
$C_{Suzaku/HXD}$ & $1.33\pm0.09$ & $1.83\pm0.13$\\
$C_{\rm \comptel}$ & $0.34\pm0.21$ & $0.29^{+0.94}_{-0.25}$ \\
$\chi^{2}/{\rm d.o.f.}$ & 2327/2501 & 2954/2941\\
$F_{\rm bol}$\tablefootmark{a} ($10^{-11}$ erg s$^{-1}$ cm$^{-2}$) & 6.77$\pm$0.36 & 2.77$\pm$0.18\\
\hline  
\end{tabular}  
\tablefoot{Best parameters determined from the fits to the average
broad-band spectrum of \ls\  performed with the \xmm/\suzaku/\nustar/\integ/\comptel\ data,
using the absorbed two-segment broken power-law (i.e. with one break energy) model. The multiplication factor for all instruments are provided.\\
\tablefoottext{a}{Unabsorbed flux in the 0.5--$2\times10^4$ keV energy range.}}
\label{tab:table3} 
\end{table} 

\section{Discussion and Conclusions} 
\label{sec:sdiscussion} 

We analyzed a vast amount of \isgri\ data for \ls\  in order to investigate its $\sim$ 100~keV emission 
for the first time and discovered a spectral break at $\sim$ 50~keV, which should be accounted for within future quantitative models of high-energy emission from the binary.

Although the nature of the compact object in \ls\ is not
yet  reliably established \citep[see][]{2000Sci...288.2340P,2005A&A...442....1A,2015MNRAS.451...59M}, a number of
studies in the past decade have successfully reproduced the main features of its light curves and
spectra assuming the presence of a neutron star \cite[see,
e.g.,][]{2013A&A...551A..17Z,Takata14,2015MNRAS.451...59M,2015A&A...581A..27D}. The interaction of a pulsar wind with the fast wind of a massive star LS 2883 is established
in the case of another well-known gamma-ray binary system associated with a radio pulsar PSR
B1259-63 orbiting around a Be star with a period of $\sim$ 3.4 yr
\citep[][]{1992ApJ...387L..37J,1999MNRAS.304..359C,2005A&A...442....1A,2011ApJ...736L..11A,2020MNRAS.497..648C}.
Although the systems have different periods and wind geometries, which certainly should be accounted for in their models \citep[see e.g.][]{2013A&ARv..21...64D}, they likely have some similarities in the multiwavelength spectra. Interestingly, PSR B1259-63 also shows hard X-ray indices $\Gamma<$~1.5 at different orbital phases \citep{2006MNRAS.367.1201C,2020MNRAS.497..648C}.

Most of the models assuming that the compact object in \ls\ is a neutron star consider the observed X-ray emission as the synchrotron radiation of the pulsar wind leptons and the VHE emission to be in phase with the X-rays and MeV gamma-rays as the IC scattering of the stellar photons.
The VHE photon index $\Gamma_{TeV} = 1.85$ observed in the INFC phase may be produced in the Klein-Nishina regime in IC emission of leptons, producing the synchrotron X-rays with $\Gamma \sim \Gamma_{TeV} / 2 \approx 0.9$ at tens to hundreds of keV. Thus, hard X-ray indices in the \isgri\ band may be consistent with the VHE index for the INFC phase.
Such hard emission spectra imply  hard energy distributions of leptons are produced inside the source. The X-ray spectra of \ls\ presented in Tables \ref{tab:table2} and \ref{tab:table3} and shown in Fig.~\ref{fig:spec_eeuf} require a specific particle distribution  in the synchrotron model. Such a distribution may be a product of a broken power law particle spectrum with a hard high-energy component.
While the power-law index $\Gamma = \left(s+1\right)/2 \sim 1.6$ assumes a particle distribution
$f\left(E\right) \propto E^{-s}$ with $s=2.2$, which is quite typical for spectra energized at the termination shocks of pulsar winds, a much harder index 
$s \sim$~1 expected above the break energy up to some limit $E=E_{max}$ would be explained by an acceleration process of some kind.
Because of the severe energy losses in strong magnetic and photon fields of an O-class star, the maximal particle energy achieved in this process giving the $E_{max}$ value may be defined by matching the acceleration timescale
with the energy loss timescale \citep[see, e.g.,][]{2006A&A...456..801D}.  This acceleration process might be less efficient and provide lower $E_{max}$ during the SUPC phase where  a much softer VHE power-law index is observed. Also, the Doppler effect may boost the emission from the acceleration site in the INFC and be insignificant or may even dampen the emission in the SUPC. The acceleration site also may be simply obscured during the superior conjuction. 

Modeling of particle acceleration in a complicated structure of interacting stellar wind and relativistic pulsar wind requires accurate simulation of flow dynamics. To simulate the wind interaction, both must be properly  parametrized.
In a recent work, \citet{bosch-ramon21} modelled gamma-ray emission as Compton upscattering
of the O-star photons by the accelerated leptons of the putative cold pulsar wind in \ls\ near
the periastron of the binary. A comparison of the modelled spectra with the observed gamma-ray
emission from \ls\ led \citet{bosch-ramon21} to the conclusion that, within the considered scenario,
a strongly magnetized cold wind is favored over a weakly magnetized strongly anisotropic
wind. Constraints on the pulsar wind magnetization and the angular dependence of the carried
momentum flux are very important for a correct modeling of the pulsar wind dynamics \citep[see, e.g.,][]{2019JPhCS1400b2027P,2020JPhCS1697a2022P}.
Despite significant progress in the fluid-type simulations of the structure of gamma-ray
binaries \citep[see, e.g.,][]{2015A&A...577A..89B,2018MNRAS.479.1320B}, modeling of the phase-resolved multi-wavelength spectra for \ls\ is a challenging problem \citep[see
e.g.,][]{2015A&A...575A.112D}. Recently,  
\citet{2020A&A...641A..84M}  used semi-analytical modeling to reproduce X-ray
{\sl Suzaku},  HE {\sl Fermi} fluxes, and VHE {\sl H.E.S.S.} data for SUPC. 
Later, \citet{reimer_ea21} developed a model of wind interaction in \ls\ and computed
a synthetic spectrum of the synchrotron and IC emission of the binary for various parameters, such as system inclination. However, the available models face difficulties in explaining the hard X-rays and  the MeV flux, as well as the VHE flux in the INFC phase.
Matching up the X-ray, MeV, and VHE data still seems to be an intricate problem \citep[see,
e.g.,][]{2015A&A...581A..27D} that has not  yet been solved. Modeling of the HE emission, which possibly contains a number of components \citep[see, e.g.,][]{Takata14}, is also a complicated problem. Recently, \citet{Yoneda2021} argued for the presence of two emission components in the GeV band, including a component which does not
depend on the orbital phase.

Shock acceleration is a widely used
mechanism of particle energization in mildly magnetized flows produced by winds in massive
star--pulsar binary systems. Diffusive acceleration by a single shock 
produces power-law test particle spectra of indices $s \gsim 2$ both at relativistic and
nonrelativistic shocks \citep[][]{BE87,B12}. The corresponding synchrotron photon spectra
show the power-law photon distribution with photon indices $\Gamma = \left(s+1\right) / 2 >
1.5$.
As discussed above, the X-ray spectra of \ls\ presented in Tables \ref{tab:table2} and \ref{tab:table3} could require a specific distribution of accelerated particles, which becomes harder at higher energies. Particle spectra
with a piece-wise power-law distribution, a low-energy branch of a spectral index
$s_1 \sim$ 2, and a high-energy branch of an index   
$s_2 \sim$ 1 just before 
$E_{max}$ were obtained in models of colliding
shock flows \citep{BSPWN_2017,2019ApJ...876L...8B}. These that can be expected in the wind collision systems that may model \ls.

Fermi I type acceleration in the colliding shock flows is likely to operate in the case of supernova shock collision with a strong
stellar wind \citep[][]{BEGO2015MNRAS}, where this mechanism can accelerate protons
well above PeV energies, as well as in colliding wind binaries \citep[][]{Grimaldo19,Pittard21}. Pulsars moving supersonically relative to the surrounding medium can create
bow-shock pulsar wind nebulae (PWNe) \citep{Reyn2017,BSPWN_2017}. The colliding shock flow
mechanism can build up a spectrum of accelerated $e^{\pm}$  in the bow-shock PWNe with the "turn-up" shape, where the high-energy branch has an index of $s_2 \sim$ 1 \citep{BSPWN_2017}. This
mechanism can explain the spatially resolved hard 0.5--8 keV X-ray indices $\Gamma <$~1.5
observed by {\sl Chandra} in the bow-shock-type Geminga PWN \citep{2017ApJ...835...66P} and in
Vela PWN  \citep[see, e.g.,][]{KP08}. The kinetic modeling presented by \cite{BSPWN_2017}
can be used to explain the hard X-ray indices observed in the latter case. The modeling also
predicts spatial variations of X-ray photon indices across the nebula with relatively soft
spectra ($\Gamma \sim$~1.6) at the pulsar wind termination shock and in the nebular tail and those
with hard index  $\Gamma \sim 1$ in the bow-shock region in coincidence with the distribution of  photon
indices observed in Geminga.

The eccentricity of
the orbit of the compact object  in \ls spatially rescales the colliding flow zone and provides different magnetic and seed-photon field intensities at the acceleration site during the INFC and SUPC phases, hence
allowing higher $E_{max}$ for the INFC phase.
As such, the difference in the VHE indices may be explained by e$^{\pm}$ pairs accelerating
to higher $E_{max}$ in the colliding flows during the INFC phase. We note that the {\sl COMPTEL}
data taken in the MeV range also show harder indices for the INFC \citep{Collmar14}. 

Thus, Fermi I type acceleration in the colliding flows may in principle be responsible for the hard spectral component discovered by \isgri\ above $\sim$ 50 keV and produce spectra consistent with the known observational results for other spectral bands.
Nevertheless, the different orientation of the colliding flow zone, the Doppler boosting effects,
and the spatially separated populations of accelerated particles with different spectral slopes may
also play a significant role in the production of the multi-wavelength  spectra discussed above.  Sensitive observations with the next generation of MeV-range detectors, such as  {\sl eASTROGAM} \cite[e.g.,][]{2017ExA....44...25D}, are needed to further quantify the characteristics of the observed spectral hardening in \ls, with important implications for the models of this fascinating object. 

\begin{acknowledgements} 
We thank an anonymous referee for the constructive comments that helped us improving the paper. A.M.B., A.M.K., and A.E.P. acknowledge support from RSF grant 21-72-20020. Some of the modeling was performed at the Joint Supercomputer Center JSCC RAS and at the "Tornado" subsystem of the St. Petersburg Polytechnic University supercomputing center. ZL thanks the
International Space Science Institute in Bern for the hospitality.  
ZL was supported by National Natural Science Foundation of China  (U1938107, U1838111). 
\end{acknowledgements} 
 
\bibliographystyle{aa}
\bibliography{references0228} 

\end{document}